\documentclass{svjour3}                     
\smartqed  
\usepackage{graphicx}
\usepackage{amsmath}
\journalname{General Relativity and Gravitation}

\def\n{\nonumber}
\def\p{\partial}
\def\be{\begin{equation}}
\def\ee{\end{equation}}
\def\d{{\rm d}}
\def\ba{\begin{eqnarray}}
\def\ea{\end{eqnarray}}
\def\ve{\varepsilon}

\begin{document}

\title{New shear-free relativistic models with heat flow}


\author{A. M. Msomi \and K. S. Govinder    \and  S. D. Maharaj}


\institute{
A. M. Msomi \and K. S. Govinder       \and  S. D. Maharaj \at
Astrophysics and Cosmology Research Unit,
 School of Mathematical Sciences,
 University of KwaZulu-Natal, Private Bag X54001,Durban 4000, South Africa
\\ \\
\email{maharaj@ukzn.ac.za} \\ \\
A. M. Msomi \at
Permanent address: Department of Mathematical Sciences,
Mangosuthu University of Technology,  P O Box 12363, Jacobs 4026, South Africa}

\date{Received: date / Accepted: date}

\maketitle

\begin{abstract}
We study shear-free spherically symmetric relativistic models with heat flow. Our analysis
is based on Lie's theory of extended groups
applied to the governing field equations.  In particular, we generate a five-parameter family of
transformations  which enables us to map existing solutions to new solutions. All known solutions
of Einstein equations with heat flow can therefore produce infinite families of new solutions.
In addition, we provide two new classes of solutions utilising the Lie  infinitesimal generators.
These solutions generate an infinite class of solutions given any one of the two unknown metric functions.


\end{abstract}

\section{Introduction} \label{sec:1}

In this paper, we consider  spherically symmetric radiating spacetimes with vanishing shear which are important in
relativistic astrophysics, radiating stars and cosmology. The assumption that the shear vanishes in a spherically
symmetric spacetime, in the presence of nonvanishing heat flux, is often made to describe the dynamics of
cosmological models. The importance of relativistic heat conducting fluids in modeling inhomogeneous processes,
 such as galaxy formation and evolution of perturbations, has been pointed out by Krasinski \cite{krasinski}. Some of the
  early exact solutions in the presence of heat flow were given by Bergmann \cite{berg}, Maiti \cite{maiti} and Modak \cite{modak}.
  Deng \cite{deng} provided a general method of generating solutions to the Einstein field equations which contains
  most previously known exact solutions. Heat conducting exact solutions are necessary to generate temperature
   profiles in dissipative processes by integrating the heat transport equation as shown by Triginer and Pavon \cite{triginer}. Bulk viscosity
    with heat flow affects the dynamics of inhomogeneous cosmological models as shown by Deng and
    Mannheim \cite{deng1}. Recently Banerjee and Chatterjee \cite{banerjee} and Banerjee {\it et al} \cite{banerjee1} have investigated
    heat conducting fluids in higher dimensional cosmological models when considering spherical collapse, the
    appearance of singularities and the formation of horizons. The role of heat flow in gravitational dynamics and
    perturbations in the framework of brane world cosmological models has been highlighted by
    Davidson and Gurwich \cite{davidson} and Maartens and Koyama \cite{maartens}.

The presence of heat flux is necessary for a proper and complete description of radiating relativistic stars. The
result of Santos  {\it et al} \cite{santos}, in one of the first complete relativistic radiating models,
 indicates that the interior spacetime should contain a nonzero heat flux so that the
matching at the boundary to the exterior Vaidya spacetime is possible. Models containing heat flow in
astrophysics have been applied to problems in the gravitational collapse, black hole physics, formation of
singularities and particle production at the stellar surface in four and higher dimensions.
The study of Chang {\it et al} \cite{chang} showed that the process of gravitational collapse of a spherical star with heat flow
may serve as a possible energy mechanism for gamma-ray bursts.

Herrera {\it et al} \cite{herrera}, Maharaj and Govender \cite{maharaj}, and Misthry {\it et al} \cite{misthry} have shown that relativistic radiating
stars are useful in the investigation of the cosmic censorship hypothesis and in describing collapse with vanishing
tidal forces. Wagh {\it et al} \cite{wagh} presented solutions to the Einstein field equations for a shear-free spherically
symmetric spacetime, with radial heat flux by choosing a barotropic equation of state. For particles in geodesic
motion a general analytic treatment is possible and solutions are obtainable in terms of elementary
and special functions  as demonstrated by Thirukkarash and Maharaj \cite{thirukk}. Herrera {\it et al} \cite{herrera1} found analytical
solutions to the field equations for radiating collapsing spheres in the diffusion approximation.
These authors demonstrated that the thermal evolution of the collapsing sphere which can be modeled
 in causal thermodynamics requires heat flow. Note that stellar models with shear are difficult
 to analyse but particular exact solutions have been found by Naidu {\it et al} \cite{naidu} and Rajah and Maharaj \cite{rajah}.

Shear-free fluids are also essential in modeling inhomogeneous cosmological processes. Krasinski \cite{krasinski}
points out the need for radiating models in the formation of structure, evolution of voids, the study of singularities
and cosmic censorship. Heat conducting fluids are important in cosmological models in higher dimensions and
permits collapse without the appearance of an event horizon; this aspect has been studied by Banerjee and Chatterjee \cite{banerjee}. In brane world
models the presence of heat flow allows for more general behaviour than in standard general relativity,
the analogue of the Oppenheimer-Snyder model of a collapsing dust permits a radiating
brane which was proved by Govender and Dadhich \cite{mgov}.

In this paper we intend to analyse the pivotal equation previously studied by Deng \cite{deng}. He developed a
general (though {\it ad hoc}) method to generate solutions and obtained a new class of solutions which
included various known special cases (see Sect.~\ref{sec:2}.).  We adopt a systematic approach (using Lie theory) to
generalise known solutions and generate new solutions of the same equation. The basic features of
Lie symmetry analysis are outlined in Sect.~\ref{sec:3}. In Sect.~\ref{sec:4}, we extend the Deng \cite{deng} known solutions to
find new solutions to the fundamental equation utilising Lie theory.  In Sect.~\ref{sec:5}, we systematically study other
group invariant solutions admitted by the fundamental equation. For most of the symmetries, we obtain
either an implicit solution or we can reduce the governing equations to a Riccati equation which is difficult
to solve in general (though particular solutions can always be found). There are two cases in which we find
new exact solutions regardless of the complexity of the generating function chosen.
Some brief concluding remarks are made in Sect.~\ref{sec:6}.

\section{Radiating Model} \label{sec:2}

Due to the requirements of spherical symmetry and the shear-free condition, the line element can be written as
\be
ds^2 = -D^2dt^2+\frac{1}{V^{2}}\left[dr^2+r^2(d\theta^2+\sin^2 \theta d\phi^2)\right]
\ee
where $D$ and $V$ are functions of $t$ and $r$. In the study of solutions of the Einstein equation with heat flux, Deng \cite{deng}
 studied a shear-free spherically symmetric cosmological model where he considered the energy-momentum tensor as
\be
T_{\mu\nu}= (\rho+p)U_{\mu}U_{\nu}+pg_{\mu\nu}+q_{\mu}U_{\nu}+q_{\nu}U_{\mu}
\ee
where $U_{\mu}$, $\rho$, $p$  are the four-velocity of the fluid, energy density and pressure, respectively, and  $q_{\mu}$ is the heat flux. The Einstein field equations are given by
\begin{subequations}
\ba
\rho &=& \frac{3V_{t}^2}{D^2V^2}+V^2\left[\frac{2V_{rr}}{V}-\frac{3V_{r}^2}{V^2}+\frac{4V_{r}}{rV}\right]\label{3a}\\
p &=& \frac{1}{D^2}\left[\frac{2V_{tt}}{V}-\frac{5V_{t}^2}{V^2}-\frac{2D_{t}V_{t}}{DV}\right]\n \\&& \mbox{}+V^2\left[\frac{V_{r}^2}{V^2}
-\frac{2D_{r}V_{r}}{DV}+\frac{2D_{r}}{rD}-\frac{2V_{r}}{rV}\right]\label{p1}\\
p &=& \frac{2V_{tt}}{D^2V}-\frac{5V_{t}^2}{D^2V^2}-\frac{2D_{t}V_{t}}{D^3V}+\frac{D_{r}V^2}{rD}\n \\&& \mbox{}-\frac{VV_{r}}{r}+\frac{D_{rr}V^2}{D}
+V_{r}^2-VV_{rr}\label{p2}\\
q &=& -2V^2\left[\frac{V_{tr}}{DV}-\frac{V_{t}V_{r}}{DV^2}-\frac{D_{r}V_{t}}{D^2V}\right]\label{3d}
\ea
\end{subequations}
Equations $(\ref{p1})$ and $(\ref{p2})$ together imply
\be
VD_{uu}+2D_{u}V_{u}-DV_{uu} = 0 \label{ben}
\ee
which is the condition of pressure isotropy with $u =r^2$. Glass \cite{glass} and Bergmann \cite{berg}, also discovered that in the
comoving system, Einstein field equations generate  the pressure isotropy condition given by the equation (\ref{ben}), which is the master equation for the system (\ref{3a})-(\ref{3d}).

A number of authors have obtained various solutions to (\ref{ben}), among which is the conformally flat class
\begin{subequations}
\ba
D &=& \frac{c(t)u+d(t)}{a(t)u+b(t)}\label{ben1}\\
V &=& a(t)u+b(t)\label{ben2}
\ea
\end{subequations}
where $a, b, c,$ and $d$ are arbitrary functions of $t$. Sanyal \cite{sanyal} and Modak \cite{modak} obtained this
class along with other solutions, while Bergmann \cite{berg} and Maiti \cite{maiti}  obtained special cases of the class.

A method of generating more general solutions to the master equation (\ref{ben}) has been developed by
Deng \cite{deng} who found solutions when simple forms of $V$ or $D$ are chosen. In finding solutions, Deng \cite{deng}
considered the master equation as an ordinary differential equation with respect to $u$. He treated (\ref{ben}) as a linear equation of $V$ if $D$ is a known function and vice versa.  His approach was as follows:
\begin{itemize}
\item Choose a simple function $D = D_{1}$ and find the most general solution $V = V_{1}$ of (\ref{ben}).
\item Take $V = V_{1}$ and find the most general solution $D = D_{2}$ obeying (\ref{ben}).
\item Take $D = D_{2}$ and find the most general solution $V = V_{2}$.
\end{itemize}
This procedure can be continued indefinitely. By alternating between $D$ and $V$, this process can go on forever generating
 infinite series of solutions expressed in terms of integrals. This is a powerful method as all known solutions can be generated using this algorithm

In this paper, we show the Deng \cite{deng} general method of generating solutions, that gives a general class of
solutions which include (\ref{ben1})-(\ref{ben2}) as special cases, may be extended by a simple invariant transformations.
In addition, we reduce the order of (\ref{ben}) via Lie analysis to obtain new solutions not obtainable via the Deng approach.

\section{Lie analysis} \label{sec:3}

The basic feature of  Lie analysis for a system of ordinary differential equations in two dependent variables requires the determination of the
one-parameter ($\ve$) Lie group of transformations
\begin{subequations}
\ba
\bar{u} &=& f(u,V,D,\ve)\label{2.2a}\\
\bar{V} &=& g(u,V,D,\ve)\\
\bar{D} &=& h(u,V,D,\ve) \label{2.2a1}
\ea
\end{subequations}
that leaves the solution set of the system invariant.
It is difficult to calculate these transformations directly.  As a result, one tends to look for the infinitesimal form of the transformations, {\it viz.}
\begin{subequations}
\ba
\bar{u} &=& u + \ve \xi(u,V,D) + O(\ve^2)\\
\bar{V} &=& V + \ve \eta(u,V,D) + O(\ve^2)\\
\bar{D} &=& D +\ve \zeta(u,V,D) + O(\ve^2). \label{2.2}
\ea
\end{subequations}
This form of the transformations can be obtained once we obtain their  ``generator'' \be  G = \xi \frac{\p\ }{\p u} + \eta \frac{\p\ }{\p V}+ \zeta \frac{\p }{\p D} \label{2.4} \ee
(also called a symmetry of the differential equation)
which is a set of vector fields. Having found the symmetries,  we can regain the finite (global) form of the transformation (\ref{2.2a})-(\ref{2.2a1}),  by solving
\begin{subequations}
\ba
\frac{\d \bar{u}}{\d \ve} &=& \xi(\bar{u},\bar{V}, \bar{D}) \label{bata}\\
\frac{\d \bar{V}}{\d \ve} &=& \eta(\bar{u},\bar{V}, \bar{D})\\
\frac{\d \bar{D}}{\d \ve} &=& \zeta(\bar{u},\bar{V}, \bar{D})\label{bata1}
\ea
\end{subequations}
subject to
\be \bar{u}\left|_{\ve=0}=u, \qquad \bar{V}|_{\ve=0}=V, \qquad \bar{D}\right|_{\ve=0}=D.\label{batas} \ee
(The full details can be found in a number of excellent texts such as Bluman and Kumei \cite{blum} and Olver \cite{olver}).

The determination of the generators  is a straight forward process, greatly aided by computer algebra packages (see the treatments of Dimas and Tsoubelis \cite{dimas} and Cheviakov \cite{chev}). We have found the
package {\tt PROGRAM LIE}  by Head \cite{head} to be the most useful in practice. It is quite remarkable how accomplished such an old package is; it often yields results when its modern counterparts fail.

Utilising {\tt PROGRAM LIE}, we can demonstrate that (\ref{ben}) admits the following Lie point symmetries/vector fields:
\begin{subequations}
\ba
G_{1} &=& \frac{\p}{\p u}\label{bobos}\\
G_{2} &=& u\frac{\p}{\p u}\\
G_{3} &=& D\frac{\p}{\p D}\\
G_{4} &=& V\frac{\p}{\p V}\\
G_{5} &=& u^2\frac{\p}{\p u}+uV\frac{\p}{\p V}\label{bobobo}
\ea
\end{subequations}

At this stage, it is usual to use these symmetries to reduce the order of the equation in the hope of finding solutions.  Before
we proceed with this approach, we show how the finite form of the transformations
generated by these symmetries can generate new solutions from known solutions.

\section{Extending known solutions} \label{sec:4}

Having found the symmetries of (\ref{ben}) we know that they generate transformations of the form (\ref{2.2a})-(\ref{2.2a1}) that leaves (\ref{ben}) invariant.

We illustrate the approach using the infinitesmal generator $G_{1}$. First we observe that
\ba
\xi &=& 1, \qquad \eta = 0, \qquad \zeta = 0
\ea
We solve equations (\ref{bata})-(\ref{bata1}), subject to (\ref{batas}), to obtain
\begin{subequations}
\ba
\bar{u} &=& u + a_{1} \label{mh1}\\
\bar{V} &=& V\\
\bar{D} &=& D \label{mh3}
\ea
\end{subequations}
This means that (\ref{mh1})-\ref{mh3}maps equation (\ref{ben}) to the form
\be
\bar{V}\bar{D}_{\bar{u}\bar{u}}+2\bar{D}_{\bar{u}}\bar{V}_{\bar{u}}-\bar{D}\bar{V}_{\bar{u}\bar{u}} = 0. \label{neweq}
\ee
As a result, any existing solution to equation (\ref{ben}) can be transformed to a solution of (\ref{neweq}) (and so a solution
of (\ref{ben}) itself) by (\ref{mh1})- \ref{mh3}. Note that, usually, $a_{1}$ is an arbitrary constant. However, since $V$
and $D$ depend on $u$ and $t$ we take $a_{1}$ to be an arbitrary function of time, $a_{1}=a_{1}(t)$.

If we now take each of the remaining symmetries successively, we obtain the general transformation as follows:
\begin{subequations}
\ba
G_{1}: \bar{u}&=& a_{1}+u, \qquad\qquad \quad \bar{D} = D,\hspace{12mm}  \bar{V}=V\\
G_{2}: \bar{u}&=& e^{a_{2}}(a_{1}+u), \qquad\quad   \bar{D} = D, \hspace{12mm} \bar{V}= V\\
G_{3}: \bar{u}&=& e^{a_{2}}(a_{1}+u), \qquad \quad \bar{D} = e^{a_{3}}D, \qquad \bar{V} = V\\
G_{4}: \bar{u}&=& e^{a_{2}}(a_{1}+u), \qquad \quad \bar{D} = e^{a_{3}}D, \qquad  \bar{V} = e^{a_{4}}V\\
G_{5}: \bar{u}&=& \frac{e^{a_{2}}(a_{1}+u)}{1-a_{5}e^{a_{2}}(a_{1}+u)},  \bar{D} = e^{a_{3}}D, \qquad \bar{V} = \frac{e^{a_{4}}V}{1-a_{5}e^{a_{2}}(a_{1}+u)}
\ea
\end{subequations}
The combination of all the transformation of symmetries, therefore leads to the general relationship:
\begin{subequations}
\ba
\bar{u}&=& \frac{e^{a_{2}}(a_{1}+u)}{1-a_{5}e^{a_{2}}(a_{1}+u)}\label{mh20}\\
\bar{D} &=& e^{a_{3}}D \\
\bar{V} &=& \frac{e^{a_{4}}V}{1-a_{5}e^{a_{2}}(a_{1}+u)} \label{mh2}
\ea
\end{subequations}
where the $a_{i}$ are all arbitrary function of time.

Thus any known solution of equation (\ref{ben}) can be transformed to a new solution of equation (\ref{ben}) via (\ref{mh20})-(\ref{mh2}). For example, the particular Deng \cite{deng} solution
\ba
D &=& 1, \qquad V = \alpha(t)u+\beta(t)
\ea
is transformed to the new solution
\begin{subequations}
\ba
\bar{u}&=& \frac{e^{a_{2}}(a_{1}+u)}{1-a_{5}e^{a_{2}}(a_{1}+u)}\\
\bar{V}&=& \frac{e^{a_{4}}(\alpha(t) u+\beta(t))}{1-a_{5}e^{a_{2}}(a_{1}+u)} \\
\bar{D} &=& e^{a_{3}}\label{shoz}
\ea
\end{subequations}
All the solutions in the Deng \cite{deng} class, the conformally flat models (\ref{ben1})-(\ref{ben2}), the result listed in
Krasinski \cite{krasinski} and Stephani {\it et al} \cite{stef1} can be extended by (\ref{mh20})-(\ref{mh2}) to produce new
solutions of (\ref{ben}). Also note that all the new results that we derive in the next section can be similarly extended via (\ref{mh20})-(\ref{mh2}).

\section{New Solutions via Lie symmetries} \label{sec:5}

The usual use of symmetries of ordinary differential equations is to reduce the order of a differential equation.
For symmetries (\ref{bobos})-(\ref{bobobo}) we obtain either an implicit solution of (\ref{ben}) or we can
reduce the governing equations to a Riccati equation. Both these results are no improvement to that
of Deng \cite{deng}, {\it i.e.} we need to choose simple forms for one of the functions (either $D$ or $V$) in order to solve for
the other. However there are two cases in which we can find new solutions regardless of the complexity of the function chosen.

\subsection{The choice $D = D(V)$}

An obvious relation to consider is when one dependent variable in (\ref{ben}) is a function of the other.
In general  such an approach usually results in a more complicated equation. However, using the
Lie symmetry $G_{1}$ (which gives the same result as $G_{2}$), we can make significant progress.
For our purposes we use the partial set of invariants of
\be
G_{1} = \frac{\p}{\p u}
\ee
given by
\begin{subequations}
\ba
p &=& V \label{mh33}\\
q(p) &=& V_{u} \\
r(p) &=& D \label{mh3}
\ea
\end{subequations}
This transformation reduces equation (\ref{ben}) to
\be
q'(p)\left[r(p)-pr'(p)\right]= q(p)\left[pr''(p)+2r'(p)\right]\label{she}
\ee
which can be integrated to give
\be
q = q_{0}e^{\int{\frac{2r'+pr''}{r-r'p}}}dp
\ee
Substituting for the metric functions via (\ref{mh33})-(\ref{mh3}),  we can integrate one more time to give the solution
\be
\int\left[e^{-\int\frac{2D_{V}+VD_{VV}}{D-VD_{V}}dV}\right]dV = q_{0}u+u_{0}\label{29}
\ee
where $q_{0}$ and $u_{0}$ are arbitrary functions of time. This means that, given any function $V$
depending on $D$, we can work out $V$ explicitly from (\ref{29}). Such a relationship
 between $V$ and $D$ has not been found previously.  Note that since (\ref{ben}) is linear,
 once we obtain $V$ via (\ref{29}) we can use it to obtain the general solution of (\ref{ben}) using standard
 techniques for solving linear equations.  (In the case of $G_{2}$, we obtain exactly the same solution as (\ref{29}).)

We illustrate this method with simple examples. Using the particular Deng \cite{deng} solution $D = 1$, we evaluate (\ref{29}) to obtain
\be
V = q_{0}(t)u+u_{0}
\ee
which is exactly what Deng \cite{deng} obtained.

If we take $D = V^2$, then equation (\ref{29}) is reduced to
\be
\int V^{6}dV = q_{0}u+u_{0}
\ee
and hence
\begin{subequations}
\ba
V_{1} &=& \left(\bar{q}_{0}u+\bar{u}_{0}\right)^{1/7}\\
D &=& \left(\bar{q}_{0}u+\bar{u}_{0}\right)^{2/7}
\ea
\end{subequations}
which is a solution of (\ref{ben}). Having found $V$ from (\ref{29}) we can generate a new solution
as follows: the second independent solution of (\ref{ben}) has the form
\be
V_{2} = y\left(q_{0}u+u_{0}\right)^{1/7}
\ee
with $y$ being unknown. Then (\ref{ben}) becomes
\be
y'' - \frac{2q_{0}}{7\left(q_{0}u+u_{0}\right)} y' = 0 \label{str}
\ee
with solution
\be
y = \frac{C_{1}}{q_{0}}\left(q_{0}u+u_{0}\right)^{9/7}+C_{2}
\ee
Hence
\be
V = \frac{C_{1}}{q_{0}}\left(q_{0}u+u_{0}\right)^{10/7}+C_{2}\left(q_{0}u+u_{0}\right)^{1/7}
\ee
where $C_1$ and $C_2$ are arbitrary functions of time, is the general solution to (\ref{ben}) when $D = \left(\bar{q}_{0}u+\bar{u}_{0}\right)^{2/7}$.

\subsection{The choice $W = \frac{V}{D}$}

We also consider the ratio of $V$ to $D$ (and later $D$ to $V$) to generate a new solution.
Incidentally both ratios arise as a result of a combination of the generators $G_{3}$ and $G_{4}$ given by
\be
G_{3}+G_{4} = D\frac{\p}{\p D}+V\frac{\p}{\p V}
\ee
This symmetry combination gives rise to the invariant
\be
W = \frac{V}{D}\label{mlomos}
\ee
Then equation (\ref{ben}) is transformed by (\ref{mlomos}) to the form
\be
 -2WD_{u}^2+D^2W_{uu} = 0 \label{xulu}
 \ee
 with solution
 \be
 D =  C_1(t)\exp\left({\int\pm \frac{\sqrt{W_{uu}}}{\sqrt{2W}}du}\right) \label{swelihle}
 \ee
Given any function $W = W(u)$ we can integrate the right hand side and find a form for $D$. If we take $W = a(t)u+b(t)$, then (\ref{swelihle}) gives \be
 D = \bar{C}_1(t)
 \ee
 and
 \be V = \bar{C}_1(t) (a(t)u+b(t))
 \ee
 which  corresponds to a Deng \cite{deng} solution.

Alternatively, we could substitute the inverse of (\ref{mlomos}), {\it i.e.}
\be \widehat{W}=\frac{D}{V}
\ee into (\ref{ben}) and obtain
\be2 D'{}^2\widehat{W}^2  - 2 D^2 \widehat{W}'{}^2 + D^2\widehat{W}\widehat{W}'' =0 \ee
with solution
\be
D = C_2(t) \exp\left(\pm\int \frac{\sqrt{\widehat{W}'{}^2 - \frac12 \widehat{W}\widehat{W}''}}{\widehat{W}}\d u \right) \label{newsol} \ee
Again, given any function $W = W(u)$ we can integrate the right hand side and find a form for $D$. If we take $W = a(t)u+b(t)$ as before,
in (\ref{newsol}),  we find that
\be D_1 = C_2(t)(a(t) u + b(t)) \ee
and
\be V_1 = C_2(t) \ee
which again is essentially a solution of Deng \cite{deng}.  However, we also have
\be D_2 = \frac{\bar{C}_2(t)}{a(t) u + b(t)} \ee
and
\be V_2 = \frac{\bar{C}_2(t)}{(a(t) u + b(t))^2} \ee
satisfies (\ref{ben}), thus obtaining two different solutions from the same seed function.

 Observe that  (\ref{swelihle}) and (\ref{newsol}) will contain all solutions of  Deng \cite{deng} for appropriately
  chosen seed functions $W$ or $\widehat{W}$ which are ratios of the metric functions.  While Deng's approach
  requires simply chosen forms of either $D$ or $V$ in order to integrate (\ref{ben}), we have no such requirement.
  We are always able to reduce (\ref{ben}) to the quadratures (\ref{swelihle}) or (\ref{newsol}) regardless of the
  complexity of the seed functions, a result not obtainable via Deng's approach.

\section{Conclusion} \label{sec:6}

We have been able to extend Deng's  solutions \cite{deng} of the Einstein field equations governing  shear-free heat
 conducting fluids in general relativity.  This was accomplished by  using simple transformations based on the
 invariance properties of the equation under study {\it viz.} (\ref{ben}). In addition, motivated by the invariants
 of the symmetries admitted by (\ref{ben}) we were able to  reduce (\ref{ben}) to quadratures for {\it any} given
 seed function. This leads to three new classes of solutions
 for infinite families of  functional forms involving  $D(V)$, $W$ and $\widehat{W}$.  This is an improvement on the approach of Deng \cite{deng} who required `simple' functional
  forms for $D$ or $V$ to be chosen before (\ref{ben}) could be solved. This again promotes the use of Lie's theory of
   extended groups to analyse the Einstein field equations arising in different applications/models in general relativity.

\begin{acknowledgements}
AMM and KSG thank the National Research Foundation and the University of
KwaZulu-Natal for financial support. SDM acknowledges that this work
is based upon research supported by the South African Research Chair
Initiative of the Department of Science and Technology and the
National Research Foundation.
\end{acknowledgements}


\begin{thebibliography}{}

\bibitem{krasinski}
Krasinski, A.: Inhomogeneous Cosmological Models. Cambridge University Press, Cambridge (1997)

\bibitem{berg}
Bergmann, O.:  Phys. Lett. A {\bf 82}, 383 (1981)

\bibitem{maiti}
Maiti, S.R.: Phys. Rev. D {\bf 25},  2518 (1982)

\bibitem{modak}
Modak, B.:  J. Astrophys. Astr. {\bf 5}, 317 (1984)

\bibitem{deng}
Deng, Y.:  Gen. Relativ. Gravit.  {\bf 21}, 503 (1989)

\bibitem{triginer}
Triginer, J., Pavon D.:  Class. Quantum Grav. {\bf 12}, 689 (1995)

\bibitem{deng1}
Deng, Y., Mannheim, P.D.:  Phys. Rev. D {\bf 42},  371 (1990)

\bibitem{banerjee}
Banerjee, A., Chatterjee, S.: Astrophys. Space Sci. {\bf 299},  219 (2005)

\bibitem{banerjee1}
Banerjee, A., Debnath, U., Chakraborty, S.:  Int. J. Mod Phys. D {\bf 12}, 1255 (2003)

\bibitem{davidson}
Davidson, A., Gurwich, I.:  JCAP  {\bf 6}, 001 (2008)

\bibitem{maartens}
Maartens, R.,  Koyama, K.:  http:// www.living reviews.org/Irr-2010-5 (2010)

\bibitem{santos}
Santos, N.O., De Oliveira, A.K.G., Kolassis, C.A.: Mon. Not. R. Astron. Soc. {\bf 216}, 1001 (1985)

\bibitem{chang}
Chang, Z., Guan, C.B., Huang, C.G., Lin. X.:  Commun. Theor. Phys. {\bf 50},  271 (2008)


\bibitem{herrera}
Herrera, L., Le Denmat, G., Santos, N. O., Wang, G.:  Int. J. Mod. Phys. D {\bf 13}, 583 (2004)

\bibitem{maharaj}
Maharaj, S.D., Govender, M.: Int. J. Mod. Phys. D {\bf 14}, 667 (2004)


\bibitem{misthry}
Misthry, S.S., Maharaj, S.D., Leach, P.G.L.: Math. Meth. Appl. Sci. {\bf 31}, 363 (2008)

\bibitem{wagh}
Wagh, S.M., Govender, M., Govinder, K.S., Maharaj, S.D., Muktibodh, P.S.,  Moodley, M.: Class. Quantum Grav. {\bf 18}, 2147 (2001)

\bibitem{thirukk}
Thirukkanesh, S.,  Maharaj, S.D.: J. Math. Phys. {\bf 50}, 022502 (2009)

\bibitem{herrera1}
Herrera, L., Di Prisco, A., Ospino, L.: Phys. Rev. D {\bf 74}, 044001 (2006)

\bibitem{naidu}
Naidu, N.F., Govender, M., Govinder, K. S.: Int. J. Mod. Phys. D {\bf 15}, 1053 (2006)

\bibitem{rajah}
Rajah, S.S., Maharaj, S.D.:  J. Math. Phys. {\bf 49}, 012501 (2008)

\bibitem{mgov}
Govender, M., Dadhich, N.K.: Phys. Lett. B  {\bf 538}, 233 (2002)

\bibitem{glass}
Glass, E.N.:  J. Math. Phys. {\bf 31}, 1974 (1990)

\bibitem{sanyal}
Sanyal, A.K.: J. Math. Phys. {\bf 25}, 1975 (1984)

\bibitem{blum}
Bluman, G.W., Kumei S.K.: Symmetries and Differential Equations. Springer, New York (1989)

\bibitem{olver}
Olver, P.J.: Applications of Lie Groups to Differential Equations. Springer-Verlag, NewYork (1993)

\bibitem{dimas} Dimas, S.,  Tsoubelis, D.: Proceedings of the 10th International Conference in
 Modern Group Analysis. (Edited by Ibragimov, N.H., Sophocleous, C., Pantelis, P.A.)
 University of  Cyprus, Larnaca  (2005)

\bibitem{chev}
Cheviakov, A.F.: Comput. Phys. Comm. {\bf 176}, 48 (2007)

\bibitem{head}
Head, A.K.:  Comp. Phys. Comm. {\bf 77} , 241 (1993)

\bibitem{stef1}
Stephani, H., Kramer, D., MacCallum, M.A.H., Hoenselaers, C., Herlt, E.: Exact solutions to Einstein's field equations.
Cambridge University Press, Cambridge (2003)










\end{thebibliography}
\end{document}